\documentclass[twocolumn,showpacs]{revtex4}

\usepackage{epsfig}
\usepackage{dcolumn}
\usepackage{amsmath}

\begin{document}

\title{{\it Ab-initio} transport theory for digital ferromagnetic 
heterostructures} 

\author{Stefano~Sanvito}
\email{e-mail: ssanvito@mrl.ucsb.edu} 
\author{Nicola~A.~Hill}
\affiliation{\normalsize Materials Department, University of California,
Santa Barbara, CA 93106, USA} 

\date{\today}

\begin{abstract}
MnAs/GaAs superlattices, made
by $\delta$-doping GaAs with Mn, are known as digital ferromagnetic
heterostructures.
Here we present a theoretical density functional study of 
the electronic, magnetic and transport properties of such heterostructures. 
In the absence of intrinsic donors these systems show an half metallic density 
of states, with an exchange interaction much stronger than that of a random
alloy with the same Mn concentration. {\it Ab initio} ballistic 
transport calculations show that the carriers with energies close to the 
Fermi energy are strongly confined within a few monolayers around the MnAs plane.
This strong confinement is responsible for the large exchange coupling.
Therefore the system can be described as a two dimensional half metal with 
large conductance in the MnAs plane and small conductance in the perpendicular 
direction.
\end{abstract}

\maketitle

Diluted magnetic semiconductors (DMS) based on III-V materials 
\cite{Ohno1,Ohno2} are receiving great attention for the intriguing interplay 
between magnetic and electronic properties \cite{Ohno3} and for the large 
potential for future electronic devices \cite{Prinz}. 
Ga$_{1-x}$Mn$_x$As, which is structurally compatible with most epitaxially
grown III-V's, is the prototype of these materials.
Here the Mn ions substitute randomly for the Ga ions, providing 
both localized $S=5/2$ spins and free carriers (holes). 
The antiferromagnetic exchange coupling between the
hole spin and the Mn spin is responsible for the long range ferromagnetic
interaction, which can be described in a first approximation by the Zener model
\cite{Diet1}. One important result of the model is that the Curie temperature is
a linear function of the Mn concentration. Although the inclusion in the model
of correlation effects \cite{ddsarma} modifies this linear dependence, there is
general agreement that an increase in the Mn concentration enhances the Curie
temperature of the system.

Unfortunately the solubility limit of Mn in GaAs is rather small, since MnAs
naturally occurs in a hexagonal phase with [6]-coordinated Mn \cite{us1}.
However large Mn concentration, up to 50\% can be obtained in zincblende MnAs 
sub-monolayers embedded into GaAs to form a MnAs/GaAs superlattice \cite{Rol1}. 
These structures, called digital 
ferromagnetic heterostructures (DFH) have remarkable properties. 
First, the Curie temperature decays for increasing GaAs inter-layer
thickness, and saturates for thicknesses larger than $\sim$50 GaAs
monolayers. The value of $T_c$ at saturation depends on the Mn
concentration \cite{Rol1}. A saturation is unexpected according to the Zener
model for three dimensional systems, since the total Mn concentration in the
sample decreases with the increase of the GaAs thickness. This separation
dependence suggests that DFH's behave like planar systems, 
although to date there is no direct proof that the free carriers
responsible for the magnetic interaction are confined in the MnAs layers.
Secondly, Hall measurements in the direction parallel to the MnAs planes show 
an anomalous Hall effect for undoped samples, which disappears upon Be-doping 
\cite{gwin1,gwin2}. Large Shubnikov de Haas oscillations are found in 
doped samples, although surprisingly the charge densities
extracted from the Hall coefficient and from the Shubnikov de Haas oscillations
are different. This suggests that two different carrier types could
be present in the system.

In this letter we investigate theoretically the magnetic and transport properties
of undoped DFH. In particular we address the following questions: i) what is the
real dimensionality of the system? ii) are the carriers spin-polarized? iii)
what is the carrier distribution in the system?

\begin{figure}[hbtp]
\centerline{\epsfig{file=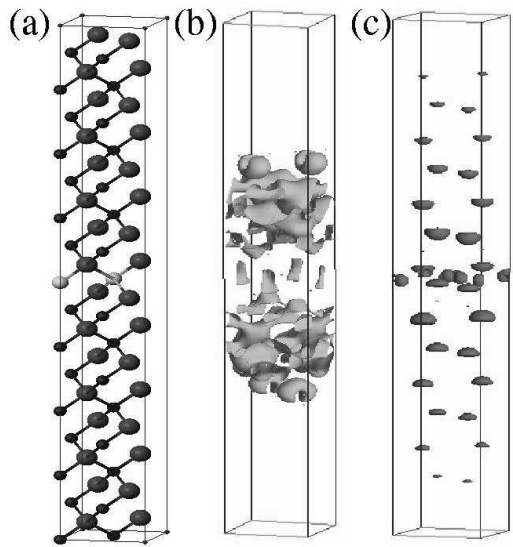,scale=0.33,angle=0}}
\caption{Charge density distribution in real space for MnAs$_1$/GaAs$_{15}$ (a)
calculated only for those states contributing to the conductance and 
energy within 0.3~eV from $E_F$: CIP (b) and CPP (c) configurations.}
\label{F4}
\end{figure}

We perform density functional theory \cite{kohn} calculations in the 
local spin density approximation. The numerical implementation 
included in the code {\sc siesta} \cite{siesta} uses pseudopotentials and a 
localized pseudoatomic orbital basis set, and allows us to investigate diluted
systems within a supercell approach. Extensive details of the method and its optimization
have already been presented \cite{us2}. Each DFH superlattice is
constructed by considering $N$ GaAs cubic cells (8 atoms in the cell) aligned
along the $z$ direction (see figure \ref{F4}a). One Ga plane is substituted 
with Mn and periodic boundary conditions are applied. 
This leads to an infinite MnAs$_1$/GaAs$_{2N-1}$ superlattice, where MnAs 
zincblende monolayers are separated by a $5.65\times N$\AA\ thick GaAs layer.

\begin{figure}[hbtp]
\centerline{\epsfig{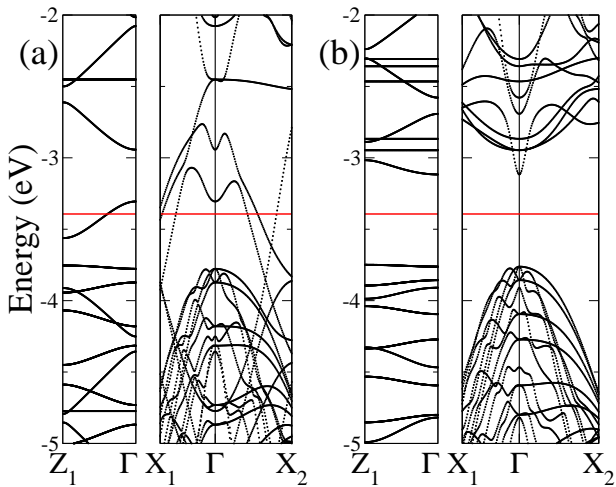}}
\caption{Band structure for a MnAs$_1$/GaAs$_{15}$ superlattice: (a) majority
and (b) minority spins. $X_1$ and $X_2$ are in the MnAs plane and
denote directions along respectively the edge and the diagonal of the cubic
supercell. $\Gamma\rightarrow Z_1$ is the direction orthogonal to the MnAs
plane. The horizontal line indicates the Fermi energy.}
\label{F1}
\end{figure}

In figure \ref{F1} we present the band structure for the case $N=8$ (with a 
45.2\AA\ thick GaAs inter-layer), for both the majority and minority spins.
The points $X_1$ and $X_2$ are in the MnAs plane and denote directions along 
the edge and the diagonal of the cubic supercell respectively. 
$\Gamma\rightarrow Z_1$ is the direction orthogonal to the MnAs plane. The band
structure of figure \ref{F1} is that of an half metal, with a gap of
approximately 0.65~eV in the minority spin band. The majority spin band presents a
strongly anisotropic dispersion. It is metallic and quite dispersive in the 
MnAs plane (directions $X_1\rightarrow\Gamma\rightarrow X_2$), while in the
direction orthogonal to the MnAs plane there is the formation of a narrow
impurity band. Since the Fermi energy $E_F$ cuts through the impurity band, 
MnAs/GaAs looks like a two dimensional half metal with small hopping between the MnAs
planes. The width of the impurity band decreases with increasing separation
between the MnAs planes (665~meV, 412~meV and 256~meV respectively for $N$=4, $N$=6
and $N$=8), which is consistent with the decrease of the overlap between
the wavefunctions centered on the MnAs planes \cite{sham}.

We calculate the strength of the magnetic coupling by calculating the 
difference $\Delta_\mathrm{FA}$ between the total energy of supercells with 
either antiferromagnetic or ferromagnetic alignment of the Mn ions \cite{us3}.
We find that the ferromagnetic interaction is much stronger than that in 
a random alloy with the same Mn concentration. 
For instance in the MnAs$_1$/GaAs$_{15}$ DFH with a Mn
concentration of $x$=0.06, $\Delta_\mathrm{FA}$ is 515~meV, whereas in the
corresponding random alloy it is only 160~meV. Moreover, $\Delta_\mathrm{FA}$ 
seems to be rather independent of the GaAs thickness for the range of thicknesses
investigated here (531~meV, 533~meV and 515~meV respectively for $N$=4, $N$=6
and $N$=8), suggesting that the carriers responsible for the ferromagnetism are 
indeed strongly confined in the MnAs planes. 

We also investigate the effect of introducing As antisites (As$_\mathrm{Ga}$) 
by considering a MnAs$_1$/GaAs$_{11}$ superlattice with half a monolayer of 
As$_\mathrm{Ga}$ close to the MnAs plane. This corresponds to one As$_\mathrm{Ga}$
for two Mn, hence according to the nominal valence, to compensation. In contrast
to what expected from the Zener model, the coupling between the Mn ions is still
ferromagnetic with $\Delta_\mathrm{FA}$=70~meV. This result is similar to that
found for the tetrahedral Ga$_1$Mn$_2$As$_2$ complex in GaAs \cite{us3},
although in that case $\Delta_\mathrm{FA}$=20~meV. We interpret the
ferromagnetism at compensation in presence of As$_\mathrm{Ga}$ as due to 
localized carriers Zener-coupled to the Mn ions \cite{us3}. The MnAs/GaAs
superlattices have stronger ferromagnetic coupling than that of the 
Ga$_1$Mn$_2$As$_2$ complex since the carrier localization close to the MnAs
plane is very strong. These results suggest that also in DFH the Curie
temperature is determined by a delicate interplay between Mn ions and intrinsic
defects.

We now move to the transport properties. Here we generalize the technique of 
reference \cite{us4} to the case of non-orthogonal tight-binding model 
with singular coupling matrices. 
The conductance is calculated in the ballistic limit with both the current 
in the Mn plane (CIP) and perpendicular to the Mn plane (CPP). 
We first extract the tight-binding 
Hamiltonian and overlap matrix giving the correct charge density. 
The matrix elements are computed by numerical integration over a real space grid
\cite{siesta}. 
The resulting tight-binding Hamiltonian comprises in principle 
an arbitrary large number of nearest neighbor interactions, which is fixed by 
the cutoff radius of the pseudoatomic orbitals. We then rewrite both the
Hamiltonian and the overlap matrix in a tridiagonal form. This procedure fixes a
``natural'' length scale. In fact we redefine the unit cell in such a way
that each unit cell is coupled only to nearest neighboring cells. Hence the problem is
reduced to an effective quasi-one dimensional problem. The scattering matrix is 
then calculated by a well established Green's function technique \cite{us4},
and the spin conductance $\Gamma_\sigma$ by using the Landauer-B\"uttiker formula \cite{but} 
and integrating over the two dimensional Brillouin zone in the plane orthogonal to
the direction of transport,
\begin{equation}
\Gamma_\sigma=\frac{e^2}{h}\sum_{k_\parallel}{\mathrm{Tr}}\; 
t_\sigma(k_\parallel) t_\sigma(k_\parallel)^\dagger\;.
\label{eq1}
\end{equation}
Here $k_\parallel$ spans the two dimensional Brillouin zone in the plane 
orthogonal to the direction of transport, and $t_\sigma(k_\parallel)$ is the 
$k_\parallel$-dependent transmission matrix for the spin $\sigma$.
We assume a two spin fluid model where there is no
mixing between majority and minority spins. It is important to point out that 
in the CPP case our supercell geometry leads to narrow bands. It is then 
crucial to perform large $k_\parallel$-point sampling. Here we consider up 
to 800 $k_\parallel$-points in the two dimensional irreducible Brillouin zone.

\begin{figure}[hbtp]
\centerline{\epsfig{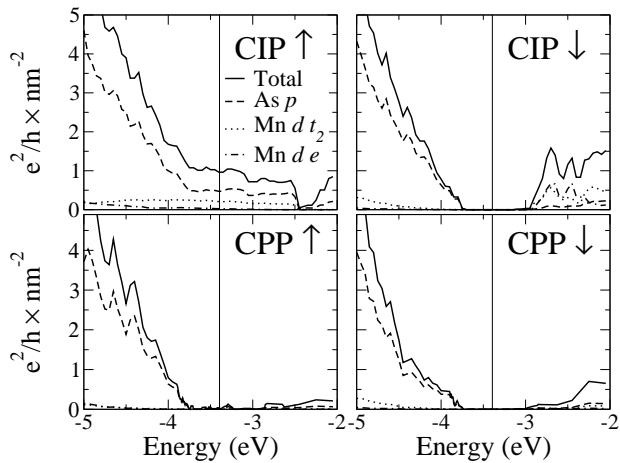}}
\caption{Total and partial conductance per unit area for a MnAs$_1$/GaAs$_{15}$ 
superlattice as a function of the position of the Fermi energy for 
majority ($\uparrow$) and minority ($\downarrow$) spin bands. 
The vertical line denotes the position of the Fermi energy for undoped samples.}
\label{F2}
\end{figure}
\begin{figure}[hbtp]
\centerline{\epsfig{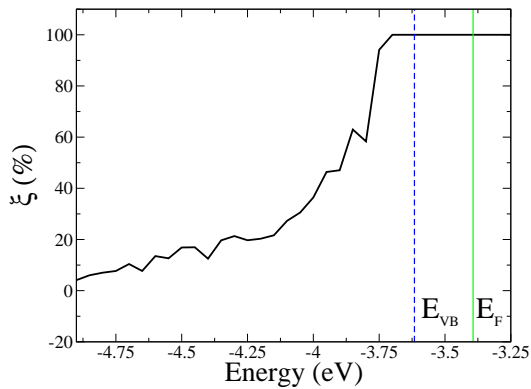}}
\caption{Spin polarization of the conductance for the CIP
configuration. The vertical lines denote the position of the Fermi energy
(solid line) and of the GaAs valence band edge (dotted line).}
\label{F3}
\end{figure}

In figure \ref{F2} we present the conductance per unit area as a function of the
position of the Fermi energy for a MnAs$_1$/GaAs$_{15}$ superlattice for both
the CIP and CPP directions and both spins. 
We also project the conductance onto the atomic orbital basis set in order to 
determine the orbital character of the electrons carrying the current \cite{us4}.
Moreover for the CIP configuration in figure \ref{F3} we show the corresponding 
spin polarization of the conductance $\xi$, which is defined as 
$\xi=\frac{\Gamma^\uparrow-\Gamma^\downarrow}
{\Gamma^\uparrow+\Gamma^\downarrow}$.
The conductance shows an half metallic behavior for both the CIP and CPP
directions. This means that there is a
100\% spin polarization of the conductance for undoped
samples ($E_F$ is in the gap for the minority spin). 
In figure \ref{F3} we also show
the position of the GaAs valence band edge, which is obtained by aligning the lower
lying As $s$ levels of bulk GaAs and the MnAs$_1$/GaAs$_{15}$ superlattice. 
It is easy to note that $\xi$ is 100\% for energies close to the GaAs valence band 
edge and that the polarization persists well within the valence band. For this 
reason DFH appear as good candidates for spin injection into $p$-doped GaAs.

If we now analyze closely the conductance of the majority spin band we notice
that it is strongly anisotropic, as expected from
the bandstructure of figure \ref{F1}. In the CIP direction the conductance as a
function of the position of the Fermi energy is rather flat with an orbital 
contribution mainly from the As $p$ orbital. 
There is also a significant contribution (roughly 
20\% of the total conductance) coming from the Mn $d$ $t_2$ orbitals,
suggesting that the carriers in undoped samples are confined close
to the MnAs planes. It is also important to point out that in $k$-space the 
contributions to the conductance are rather uniformly distributed over 
the two dimensional Brillouin zone in the plane orthogonal to the transport 
direction. These are strong indications that MnAs/GaAs is a good isotropic 
metal in the MnAs planes.

For the CPP direction the situation is rather different. The conductance as a
function of the position of $E_F$ shows clear one dimensional features and for 
undoped samples it is solely given by charge in the impurity band.
The orbital content of the impurity band is mainly As $p$ with small
contributions from Mn $d$ $t_2$. The conductance at $E_F$ comes from few $k$-points
around the $\Gamma$ point. In the case of carriers with parabolic dispersion 
this happens for scattering at a potential barrier, when the carrier energy is 
very close to the top of the barrier. In fact carriers
with largest kinetic energy in the direction orthogonal to the barrier (small 
$k$ in the parallel direction) have largest transmission amplitude.
This situation is similar to that of tunneling junctions and so we describe 
the transport as tunneling-like.

In summary, from ballistic transport calculations, MnAs/GaAs appears to be a 
two dimensional half metal in the MnAs plane, with tunneling-like conductance 
between the MnAs planes. 
In order to better understand the spatial arrangement of the current in
figure \ref{F4} we present the charge density distribution in real space,
$\rho({\bf r})$, calculated only for those states contributing to the 
conductance and within a 0.6~eV range around $E_F$
($E_F-0.3\mathrm{eV}<E<E_F+0.3\mathrm{eV}$). We also notice that the individual
contributions from occupied ($E_F-0.3\mathrm{eV}<E<E_F$) and 
unoccupied ($E_F<E<E_F+0.3\mathrm{eV}$) states are very similar.

The figure confirms that the current in the CIP case is distributed mainly in
a narrow region around the MnAs planes, with small spillage outside.
Planar averaging \cite{Bald} of the charge density of figure \ref{F4} shows that
the charge density reaches $\sim$0.1 of its MnAs plane value within 
only three GaAs monolayers from the MnAs plane. 
In contrast, the CPP current is mainly located at the Mn
plane with small contributions from the GaAs layers. This means that
carriers are strongly confined in the MnAs plane and the perpendicular transport 
is via hopping between the planes.

\begin{figure}[hbtp]
\centerline{\epsfig{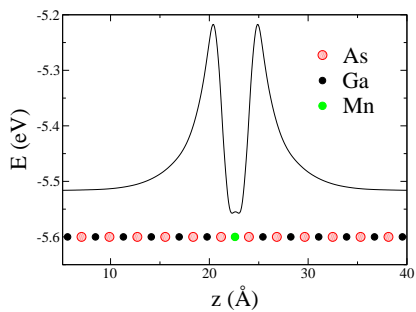}}
\caption{Macroscopic average of the Hartree potential for
MnAs$_1$/GaAs$_{15}$.}
\label{F5}
\end{figure}

Finally in figure \ref{F5} we present the macroscopic average \cite{Bald} of
the Hartree potential. The macroscopic average is obtained by first taking a
planar average and then by averaging the result over the GaAs lattice spacing.
This does not correspond exactly to the total potential 
which the quasiparticles feel in the structure, but it gives a good 
indication of the latter \cite{Bald}.
The potential profile shows a deep well in the MnAs plane separated from a flat
region by two barriers located one monolayer from the well. The presence of the 
barrier suggests that $n$-type doping in the GaAs region will only weakly affect 
the MnAs planes, since the extra electrons need to tunnel through a large barrier to
access the MnAs planes. 

In conclusion we have studied with {\it ab initio} methods the magnetic and 
transport properties of GaAs/MnAs digital ferromagnetic heterostructures. We
have shown that they are two dimensional half metallic systems with large 
metallic conductance in the MnAs plane, and small tunneling-like conductance 
perpendicular to the MnAs plane. 
We have shown that the electronic charge at the Fermi energy is mainly 
distributed within a few monolayers around the MnAs plane and that the Hartree 
potential shows a double barrier profile. This potential profile can sustain
the presence of a two carrier population for DFH's that are 
Be-doped in the GaAs region.

We thank E.~Gwinn for stimulating discussions.
This work made use of MRL Central Facilities supported by the National Science 
Foundation under award No. DMR96-32716. This work is supported by ONR 
grant N00014-00-10557, by NSF-DMR under the grant 9973076 and by ACS PRF under 
the grant 33851-G5.

\end{document}